\documentclass[%
 reprint,
 amsmath,amssymb,
 aps,
]{revtex4-1}

\usepackage{graphicx}
\usepackage{dcolumn}
\usepackage{hyperref}
\usepackage{bm}

\newcommand*{\affmark}[1][*]{\textsuperscript{#1}}

\begin{document}


\title{Self impedance matched Hall-effect gyrators and circulators}

\author{S. Bosco\affmark[1,3]}
\email {bosco@physik.rwth-aachen.de}
\author{F. Haupt\affmark[1,3]}%
\author{D. P. DiVincenzo\affmark[1,2,3]}
 \email{d.divincenzo@fz-juelich.de}
\affiliation{
\affmark[1]Institute for Quantum Information, RWTH Aachen University,                                
  D-52056
  Aachen,                              
  Germany
}

\affiliation{
  \affmark[2]Peter Gr\"{u}nberg Institute, Theoretical Nanoelectronics,
    Forschungszentrum J\"{u}lich,
  D-52425
  J\"{u}lich,
  Germany
}

\affiliation{
\affmark[3]J\"{u}lich-Aachen Research Alliance (JARA),
    Fundamentals of Future Information Technologies,
  D-52425
  J\"{u}lich,
  Germany
}

\date{\today}

\begin{abstract}

We present a model study of an alternative implementation of a two-port Hall-effect microwave gyrator.
Our set-up involves three electrodes, one of which acts as a common ground for the others.
Based on the capacitive-coupling model of Viola and DiVincenzo, we analyze the performance of the device and we predict that ideal gyration can be achieved at specific frequencies.
Interestingly, the impedance of the three-terminal gyrator can be made arbitrarily small for certain coupling strengths, so that no auxiliary impedance matching is required.
Although the bandwidth of the device shrinks as the impedance decreases, it can be improved by reducing the magnetic field; it can be realistically increased up to $ 150 \mathrm{MHz}$ at $50\mathrm{\Omega}$ by working at  filling factor $\nu=10$.
We examine also the effects of the parasitic capacitive coupling between electrodes and we find that, although in general they strongly influence the response of device, their effect is negligible at low impedance.
Finally, we analyze an interferometric implementation of a circulator, which incorporates the gyrator in a Mach-Zender-like construction. Perfect circulation in both directions can be achieved, depending on frequency and on the details of the interferometer.

\end{abstract}

\pacs{Valid PACS appear here}
\maketitle


\section{\label{sec:introduction}Introduction}

Non-reciprocal devices such as gyrators and circulators are key components in modern microwave engineering. Circulators, for example, are crucial for reducing thermal noise and for directing signals in superconducting circuits \cite{DiCarlo, 11qubits}. 
One conventional implementation of circulators exploits the electromagnetic Faraday effect in ferromagnets \cite{Hogan1, Hogan2}. Although excellent circulation can be obtained, this set-up has a fundamental scalability limit, which sets the minimal size of the device to the order of centimeters.
Alternative, more compact designs rely on active components, typically operational amplifiers \cite{OPAMP}, whose performances are currently not guaranteed at cryogenic temperatures.

Recently, Viola and DiVincenzo (VD)  \cite{Viola-DiVincenzo} revisited the old idea of obtaining non-reciprocity by exploiting the Hall effect (HE).
Previous studies of Hall-effect gyrators focused on Ohmic contacts between the external metal electrodes and the Hall bar \cite{Wick, Girvin} and showed that the device has in this case an input resistance larger than the gyration resistance, which severely degrades the performance.
In contrast, VD considered a Hall bar \textit{capacitively} coupled to external electrodes, and predicted that it could behave as an optimal non-reciprocal device in specific ranges of frequency. 
Motivated by this work, a first experimental realization of a HE circulator was attempted by Mahoney \textit{et al.} (MEA) \cite{Reilly}. Although this experiment showed the potentiality of this novel set-up, it also pointed out a number of engineering challenges to be faced.
In particular, MEA focused on the Carlin construction of a circulator \cite{Carlin1, Carlin2, Viola-DiVincenzo}, and in this case the performance of the device is strongly affected by the high impedance mismatch between the Hall conductor and the external microwave circuit and by the parasitic coupling between the electrodes. A model study of these effects in that case is extensively discussed in \cite{Placke}.

In this work, we analyze an alternative set-up for HE gyrators and circulators that are highly insensitive to impedance mismatch and parasitics.
VD proposed a bipolarly-driven four-terminal gyrator, where each pair of opposite electrodes is driven by identical voltage signals, with opposite sign.
Here, we consider instead a three-terminal gyrator, with one terminal acting as common ground. This set-up is not only experimentally easier to implement, but we also find that for certain coupling strengths it also exhibits a very interesting self-matching property: its impedance can be made arbitrarily small, so that no additional impedance matching is required.
We predict that as the impedance decreases, on one hand unwanted effects due to parasitics become negligible, on the other hand the frequency range (bandwidth) within which the device works  shrinks.
However, we also find that the bandwidth can be improved by lowering the magnetic field, and that for realistic device parameters, it can be sufficiently large for several applications, e.g. $\approx 150\mathrm{MHz}$ at filling factor $\nu=10$ and external impedance $50\mathrm{\Omega}$. 

Finally, we analyze an alternative implementation of a circulator, which incorporates the three-terminal HE gyrator into a Mach-Zender-like interferometer.
We find that the quality and the direction of circulation are crucially related to the details of the interferometer.
In particular, to have ideal circulation  in the smallest possible set-up, i.e. when all the arms of the interferometer are very short, directional couplers with both $\pi$ and $\pi/2$ phase shift are required.
In this case, the size of the circulator is mainly limited by the size of the directional couplers, which are typically implemented on-chip in $\lambda/4$ ring structures, and that can  be miniaturized to $\lambda/N$, $N\approx 20$, by standard microwave engineering techniques \citep{Simons}.

\section{\label{sec:gyrator}3-terminal gyrator}

\subsection{\label{subsec:ideal-gyrator}Basic analysis}

\begin{figure}
\includegraphics[scale=0.4]{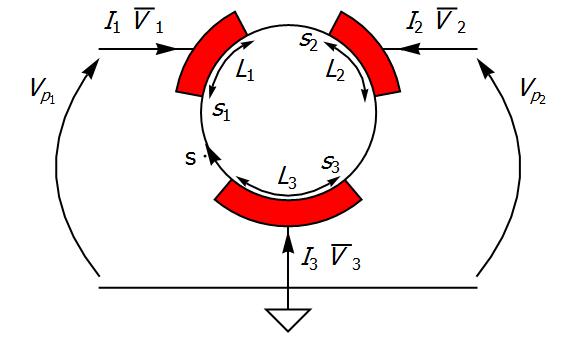}
\caption{Three-terminal gyrator. Three electrodes of length $L_i$ are capacitively coupled to a Hall bar, whose perimeter is parametrized by $s$.  The device behaves as a two-port circulator when the voltage of two electrodes is measured with respect to the last one, which acts as a common ground. The convention of currents and voltages is shown in the plot.}
\label{fig:3-terminals-gyrator}
\end{figure}

We analyze the performance of the HE gyrator composed of a two-dimensional Hall conductor capacitively coupled to three perfectly conducting electrodes, as shown in Fig. \ref{fig:3-terminals-gyrator}. 

To model the device, we follow VD and begin by computing the electric potential  $V$ inside the Hall bar. This has to satisfy the Laplace equation, 
\begin{equation}
\nabla^2V(x,y)=0,
\label{eq:Laplace}
\end{equation}
with boundary conditions accounting for the applied magnetic field and the coupling to the electrodes.
The work of VD showed that in the quantum Hall regime, i.e. when the material conductance is at a quantum Hall plateau \cite{QuantumHallGirvin}, the boundary conditions on $V$ reduce to the one-dimensional closed differential equation
\begin{equation}
    -\sigma\partial_s V(s, \omega)
    = i \omega c(s)\left( \overline{V}(\omega) - V(s, \omega) \right),
    \label{eq:capacitive-boundary}
\end{equation}
in the perimeter coordinate $s$. Here, $\overline{V}$ is the voltage-drive applied at the terminals and $\sigma=\nu e^2/h$ is the quantum Hall conductivity at filling factor $\nu$. 
The phenomenological function $c(s)$ has dimension of capacitance per unit length and it accounts for the capacitive coupling to the external metal electrodes. When the capacitors are much longer than the gap between them, one can neglect fringing fields and $c(s)$ is well approximated by the step-wise function
\begin{equation}
c(s)=\left\{
\begin{array}{ll}
      c_1 &  s_1<s<s_1+L_1 \\
      c_2 &  s_2<s<s_2+L_2 \\
      c_3 &  s_3<s<s_3+L_3 \\
      0 & \mathrm{otherwise}.
\end{array} 
\right.
\label{eq:capacitance-function}
\end{equation}
The parameters $c_i$ should incorporate both classical electrostatic-coupling and quantum-capacitance effects.  We will report separately \cite{Bosco} on a microscopic calculation, based on the RPA approximation, that gives the driven response of the 2D Hall conductor; this calculation incorporates details of the edge-conductor geometry, and the band structure and edge-confinement physics of the 2D material.  Within both hetrostructure and graphene models, we confirm that the simple approximations of the VD theory give results in agreement with RPA in many situations of interest.

A peculiar feature of the quantum Hall limit is that the boundary potential $V(s,\omega)$ depends only on the edge dimension, and thus the VD model captures the behavior of Hall bars of arbitrary shape. 
Moreover, as VD pointed out, in this case $V(s,\omega)$ fully determines the dynamics of the potential inside the material, and all the interesting parameters characterizing the device response can be computed from it.
In particular, the current in the $i$th electrode is simply \cite{Viola-DiVincenzo}
\begin{equation}
    I_i(\omega) = \sigma \left[ V(s=s_i,\omega)-V(s=s_i+L_i,\omega) \right].
    \label{eq:current-rotated}
\end{equation}

To study a two-port gyrator, we define the port voltages and currents as
\begin{subequations}
\begin{align}
V_{p_1}&=\overline{V}_1-\overline{V}_3, &&  & I_{p_1}&=I_1,\\
V_{p_2}&=\overline{V}_2-\overline{V}_3, &&  & I_{p_2}&=I_2,
\end{align}
\label{eq:ports-gyrator}
\end{subequations}
see Fig. \ref{fig:3-terminals-gyrator}.
Combining Eqs. (\ref{eq:capacitive-boundary}), (\ref{eq:current-rotated}) and (\ref{eq:ports-gyrator}), we find the port-admittance matrix $Y_p$ and, by using the standard relation \cite{Pozar}
\begin{equation}
S=-\left( Y_p+\frac{1}{Z_0}\mathcal{I}\right)^{-1}\left( Y_p-\frac{1}{Z_0}\mathcal{I}\right),
\label{eq:S-matrix-Pozar}
\end{equation}
we can evaluate the $S$ parameters of the device. In Eq. (\ref{eq:S-matrix-Pozar}), $\mathcal{I}$ is the identity matrix and $Z_0$ is the characteristic impedance of the external microwave circuit, typically $50\Omega$.

For simplicity, we now focus on a symmetric situation, with $C\equiv c_1 L_1=c_2 L_2$, and we introduce the dimensionless parameter $r\equiv c_3 L_3/C$, which characterizes the strength of the coupling to the ground electrode with respect to the others.
Substituting $Y_p$ in Eq. (\ref{eq:S-matrix-Pozar}), we find that the scattering matrix assumes the form 
\begin{equation}
S=
\left(
\begin{array}{cc}
 S_{11} & S_{21} e^{i \Omega  r} \\
 S_{21} & S_{11} \\
\end{array}
\right),
\label{eq:s-g-matrix}
\end{equation}
where $\Omega$ is the dimensionless frequency of the signal defined by
\begin{equation}
\Omega\equiv \omega C/ \sigma.
\end{equation}
The explicit expressions for $S_{11}$ and $S_{21}$ are shown in appendix \ref{appendix:S-parameter}.

From Eq. (\ref{eq:s-g-matrix}), it follows that maximal non-reciprocity is obtained at frequencies
\begin{align}
\Omega_{n} = \frac{\pi}{r}(2n+1) && \mathrm{with} && n\in \mathbb{N}^+.
\label{eq:gyration-frequency}
\end{align}
Condition (\ref{eq:gyration-frequency}) is necessary but not sufficient to achieve ideal gyration. In fact, an ideal gyrator needs to be both anti-reciprocal and reflectionless, with $S_{11}=S_{22}=0$. 
Hence, to quantitatively analyze the device, we introduce
\begin{subequations}
\begin{flalign}
\label{eq:delta-definition}
\Delta &\equiv\frac{1}{2}\lvert S_{21}-S_{12}\rvert \leq 1, \\
\label{eq:phase-gyrator}
\varphi &\equiv \arg \left(  S_{21}-S_{12} \right).
\end{flalign}
\end{subequations}
In particular, $\Delta$ characterizes the gyrator performance, since the unitarity of the $S$-matrix guarantees that  equality in Eq. (\ref{eq:delta-definition}) is attained only for ideal gyrators. The overall phase factor $\varphi$ is crucial for fixing the parameters of the interferometric implementation of a circulator.

In general, the reflection coefficients strongly depend on the impedance mismatch between the Hall bar and the external system. To characterize this we define the dimensionless parameter $\alpha\equiv 2\sigma Z_0$.

\begin{figure}
\includegraphics[scale=0.6]{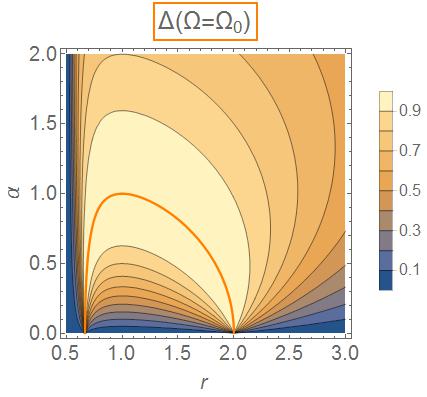}
\caption{Dependence of $\Delta$ on $r\equiv c_3 L_3/(c_1 L_1) $ and $\alpha\equiv 2\sigma Z_0$ evaluated at the first gyration frequency $\Omega_0\equiv\pi/r$.
Good gyration requires $\Delta$ near $1$.
Along the orange line, the value $\Delta=1$ is exactly attained, giving  perfect gyration.}
\label{fig:r-alpha-dep-gyrator}
\end{figure}

First, we focus on anti-reciprocal devices.
Fig.  \ref{fig:r-alpha-dep-gyrator} shows how $\Delta$ is influenced by $\alpha$ and $r$ at the first gyration frequency $\Omega=\Omega_{0}\equiv\pi/r$. 
We note that for all values of the mismatch parameter $\alpha \leq 1$, there is a value of $r$ at which perfect gyration is attained. This means that the impedance of the HE gyrator can be made arbitrary small by choosing an appropriate ratio between the coupling strength of the electrodes.
In particular, it is straightforward to show that the condition $\Delta\left(\Omega_n\right)=1$ holds when $r$ and $\alpha$ are related by the transcendental equation
\begin{equation}
\label{eq:perfect-gyrator-cond}
\cos\left(\frac{\pi}{r}(2n+1)\right)=1-\frac{1}{1-\alpha^2/2}.
\end{equation}

This striking self-matching property is a new feature of our three-terminal construction, which can be exploited to engineer compact devices not requiring external impedance matching.
From an experimental point of view, the case $\alpha\ll 1$ is of the utmost importance, since in the quantum Hall regime $\alpha\approx0.004 \nu$ at $50\Omega$. If we expand condition (\ref{eq:perfect-gyrator-cond}) for small $\alpha$, we obtain the simple formula 
\begin{align}
\label{eq:self-matching-conditions}
   r = \frac{2(2n+1)}{2(m-n)-1}+\mathcal{O}(\alpha^2) && \mathrm{with} && n, m\in \mathbb{N}^+,
\end{align}
and $m\geq n+1$.

\begin{figure}
\includegraphics[scale=0.61]{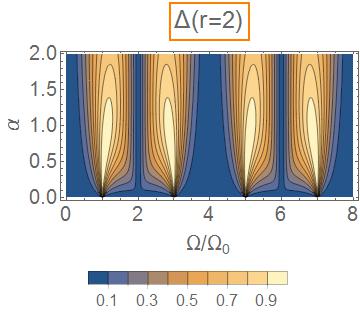}
\caption{$\Delta$ as a function of $\alpha\equiv 2\sigma Z_0$ and of the dimensionless frequency $\Omega$, normalized by the first gyration frequency $\Omega_0\equiv\pi/r$. The plot shows the result  when $r\equiv c_3 L_3/(c_1 L_1)=2$, which corresponds to the self-matched case.}
\label{fig:omega-dep-gyrator}
\end{figure}

It is now worth analyzing in more detail the frequency response of the self-matched construction. Fig. \ref{fig:omega-dep-gyrator} shows $\Delta$ as a function of $\alpha$ and $\Omega$ when $r=2$.
First of all, it is interesting to note that the response is periodic in $\Omega$, with period $2\pi$.
Moreover, although strictly speaking perfect gyration is achieved only at $\alpha\rightarrow0$ and $\Omega=\Omega_n$, excellent performance is seen in a broader range of parameters. 

\begin{figure}
\includegraphics[scale=0.55]{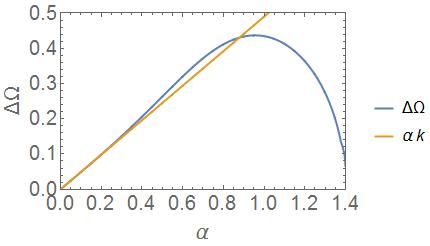}
\caption{Dependence of $\Delta\Omega$ on $\alpha$ in the self-matched case with $r=2$. The BW increases up to $\alpha=1$ and then it decreases abruptly. In the interesting impedance regime, $\alpha \lesssim 0.4$, the increase of the BW is to good approximation linear,  with slope $k\approx0.49$.}
\label{fig:BW-gyrator}
\end{figure}

To quantitatively analyze this range, we set the threshold defining a good device to $\Delta=0.9$ and we introduce the bandwidth (BW) of the device as the dimensionless frequency range $\Delta\Omega$ for which $\Delta \geq 0.9$.
Fig. \ref{fig:BW-gyrator} shows the dependence of $\Delta\Omega$ on $\alpha$. 
Increasing $\alpha$ from zero, the BW of the gyrator increases approximately linearly with slope $k\approx0.49$ until the maximal BW is obtained at $\alpha\approx 1$, which corresponds to the perfectly matched limit in \cite{Viola-DiVincenzo}.

Using a realistic capacitance value of $C=50\mathrm{fF}$, and for $r=2$, one can implement devices working at frequencies $\omega_n \approx 1.2 \nu (2n+1)\mathrm{GHz}$, with bandwidth $\Delta\omega \approx 1.5 \nu^2 \mathrm{MHz}$ at  $Z_0=50\Omega$.
Note that by increasing the filling factor $\nu$, the gyration frequency increases linearly, while the BW increases quadratically. Hence, by lowering the magnetic field, a bandwidth which is useful for several practical applications, e.g. $\approx\mathrm{150MHz}$ at $\nu=10$, can be easily reached.

\subsection{Parasitic capacitances}

\begin{figure}
\includegraphics[scale=0.4]{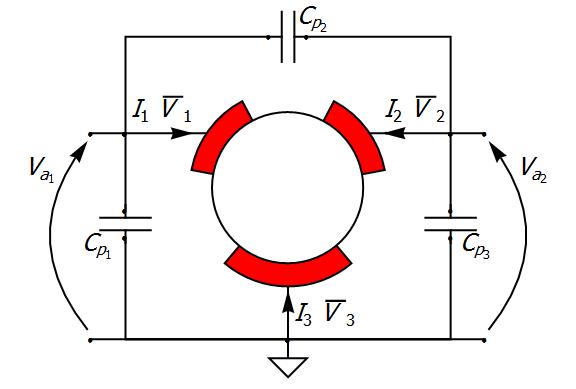}
\caption{Augmented network incorporating parasitic capacitive coupling between the three electrodes. The convention on port-voltages $V_a$ are shown.}
\label{fig:augmented-network-parasitics}
\end{figure}

Motivated by recent works, both theoretical and experimental \cite{Reilly, Placke}, we now include the effect of the parasitic capacitive coupling between electrodes, and we predict that at low enough impedance, these effects are negligible.
We consider the augmented network shown in Fig. \ref{fig:augmented-network-parasitics}. For simplicity, we assume the electrodes 1 and 2 to be placed symmetrically with respect to the third, such that two of the three parasitic capacitors are equal, $C_{p_1}=C_{p_3}$.

From straightforward circuit analysis, since the parasitic channels are in parallel with the device, the admittance matrix of the augmented network $Y_a$ is
\begin{equation}
\label{eq:Y-a}
Y_a=Y_p+Y_1+Y_2,
\end{equation}
with
\begin{subequations} 
\begin{flalign}
Y_1&=i \omega C_{p_1}\mathcal{I},\\
Y_2&=i \omega C_{p_2}
\left(
\begin{array}{rr}
  1 & -1 \\
 -1 &  1 \\
\end{array}
\right).
\end{flalign}
\end{subequations}
The scattering parameters are then readily found by using  Eq. (\ref{eq:S-matrix-Pozar}).

\begin{figure}
(a)\includegraphics[scale=0.5]{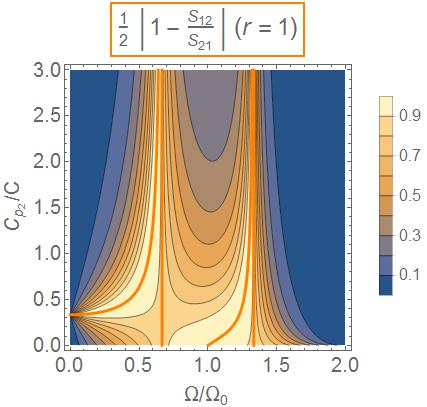}
(b)\includegraphics[scale=0.5]{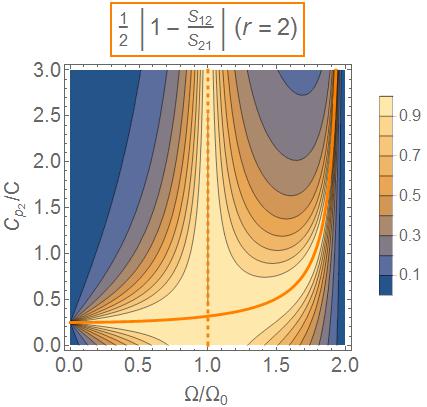}
\caption{Non-reciprocity contribution to $\Delta$ at different $r$ as a function of $C_{p_2}$ and $\Omega/\Omega_0$. We used in (a) $r=1$ and in (b) $r=2$. The ratio between the off-diagonal terms depends neither on $\alpha$ nor on $C_{p_1}$. The orange lines represent the maximally anti-reciprocal case, attained for $\lvert 1- S_{12}/S_{21}\rvert/2=1$. In particular, the dashed line corresponds to condition (\ref{eq:cond-2}), while the solid line corresponds to condition (\ref{eq:cond-4}).}
\label{fig:parasitics-r}
\end{figure}

To characterize the effect of the parasitic channels, we begin by analyzing the non-reciprocal behavior of the device. We find that  the frequency that guarantees maximal non-reciprocity is shifted depending on $C_{p_2}$.
To examine this phenomenon, we study the quantity $\lvert 1- S_{12}/S_{21}\rvert/2$, which is equal to 1 only for anti-reciprocal devices.  In Fig. \ref{fig:parasitics-r}, we show its dependence on $\Omega$ and $C_{p_2}$ for two different values of $r$. 
The behavior observed in the Figure is explained by considering that there are two different conditions that guarantee maximal anti-reciprocity:
\begin{subequations}
\begin{flalign}
r &=\frac{2 (2 n+1)}{2 (m-n)-1} \land \Omega = \Omega_n ,\  \mathrm{or}
\label{eq:cond-2} \\
\frac{C_{p_2}}{C} &= \frac{2}{\Omega} \sin ^2\left(\frac{\Omega }{2}\right) \cos \left(\frac{\Omega r }{2}\right) \csc \left(\frac{\Omega}{2} (r+2) \right),
\label{eq:cond-4}
\end{flalign}
\end{subequations}
with $\Omega_n$ being the gyration frequencies in Eq. (\ref{eq:gyration-frequency}). 

The former condition (\ref{eq:cond-2}) corresponds to the self-matching conditions for small $\alpha$ in Eq. (\ref{eq:self-matching-conditions}): as it does not depend on the parasitic capacitors, it explains the straight line in Fig. \ref{fig:parasitics-r}(b) at $\Omega=\Omega_0$. 
In contrast, the second condition (\ref{eq:cond-4}) shifts the frequency range that guarantees maximal non-reciprocity, and it governs the response for high enough gyrator impedance.

\begin{figure}
\includegraphics[scale=0.6]{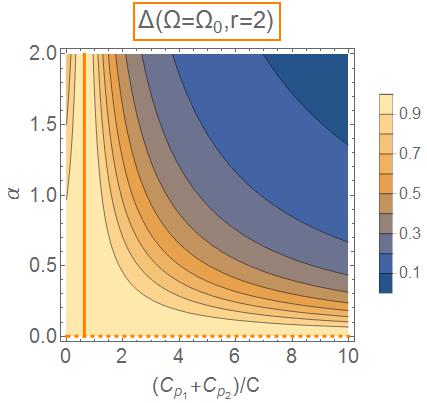}
\caption{$\Delta$ as a function of $\alpha$ and of the sum of parasitic capacitors, normalized over $C\equiv c_1 L_1$ at $r=2$ and at the first gyration frequency $\Omega=\Omega_0$.
Along the orange lines, $\Delta=1$ and perfect gyration is attained. The solid line corresponds to  Eq. (\ref{eq:parasitics-perfect-gryrat}) with  $n=m-1=0$, while the dashed line corresponds to $\alpha=0$.}
\label{fig:self-matched-parasitics}
\end{figure}

To conclude the analysis of the parasitics, we examine the reflection properties of the device focusing on the case described by Eq. (\ref{eq:cond-2}), with $n=m-1=0$. 
It turns out that in this limit $\Delta$ depends uniquely on $\alpha$ and on the sum of the two parasitic capacitors $C_{p_1}$ and $C_{p_2}$. This dependence is shown in Fig. \ref{fig:self-matched-parasitics}; it shows that our construction is barely influenced by reasonably small parasitics when the impedance mismatch is high enough. In  particular, perfect gyration is always attained in the limit $\alpha\rightarrow 0$.
Finally, it is interesting to notice that when condition (\ref{eq:cond-2}) is satisfied, $\Delta=1$ for all values of mismatch parameter $\alpha$ when
\begin{align}
\frac{C_{p_1}+C_{p_2}}{C}= \left|\pi (m- n)-\frac{\pi}{2} \right|^{-1},
\label{eq:parasitics-perfect-gryrat}
\end{align}
and $m+n$ is odd.

\section{\label{sec:circulator} Circulator}

\begin{figure}
(a)\includegraphics[scale=0.5]{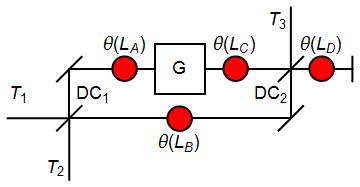}
(b)\includegraphics[scale=0.5]{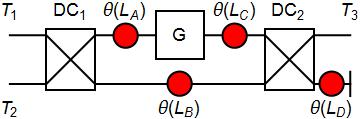}
\caption{Circulator construction in optical (a), and microwave (b) conventions. A standard Mach-Zender interferometer is modified by incorporating a gyrator (G) in one of the arms and by interrupting one of the arms after a directional couplers (DC) by a fully reflecting mirror, which in microwave engineering corresponds to either an open or  a short circuit to ground.
The phase $\theta(L_i)$ accumulated by a signal passing through the $i$th transmission line is provided by reciprocal phase-shifters. }
\label{fig:circulator-convention}
\end{figure}

We can now analyze the circulator obtained by using the three-terminal gyrator in the interferometric (Hogan) construction shown in Fig. \ref{fig:circulator-convention}.

To see which details of the interferometer play a crucial role for gyration, it is useful to study what happens for an ideal gyrator, with scattering parameters $S_{11}=S_{22}=0$ and $S_{21}=-S_{12}=e^{i \varphi}$.
The overall phase of the gyrator $\varphi$ corresponds to the quantity defined in Eq. (\ref{eq:phase-gyrator}) when the device is anti-reciprocal.
It is straightforward to find that ideal circulation imposes the condition
\begin{equation}
e^{-i 2 \theta(L_B)}=\beta e^{-i 2 (\theta(L_A)+ \theta(L_C)-\varphi)},
\label{eq:circulation-condition}
\end{equation} 
where $\theta(L_i)$ is the phase that a signal accumulates in passing through the $i$th lossless transmission line. For TEM modes of transmission lines it is simply $\theta(L_i)=2\pi L_i/\lambda$, with  $\lambda$ being the wavelength of the signal.
The parameter $\beta$ depends on the particular combination of directional couplers used: $\beta=1$ for two equal directional couplers, and $\beta=-1$ if the directional couplers are $\pi/2$ out-of-phase.
Moreover, when the condition (\ref{eq:circulation-condition}) is satisfied, the phase $\theta(L_D)$ does not affect the absolute value of the scattering parameters, although it changes the relative phase of the signals at the different electrodes.

\begin{figure}
\includegraphics[scale=0.6]{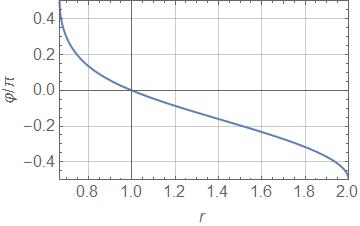}
\caption{Phase of the ideal HE gyrator. We focused on the first gyration frequency $\Omega=\Omega_0$, which guarantees that $\varphi$ as defined in equation (\ref{eq:phase-gyrator}) is the actual phase of the gyrator. The plot shows $\varphi$ when $r$ and $\alpha$ are related by the condition (\ref{eq:perfect-gyrator-cond}), which corresponds to the path described by the solid orange line in Fig. \ref{fig:r-alpha-dep-gyrator}.}
\label{fig:phase-gyrator}
\end{figure}

To have the most compact implementation of the circulator, we set $L_A=L_C=L_D=0$, and then the length $L_B$ is fixed by $\beta$ and $\varphi$ according to Eq. (\ref{eq:circulation-condition}). 
As anticipated in Sec. \ref{subsec:ideal-gyrator}, the overall phase of the gyrator plays now a fundamental role.
Fig. \ref{fig:phase-gyrator} shows how $\varphi$ varies at $\Omega=\Omega_0$ as a function of $r$ on the path defined by the ideal gyration condition (\ref{eq:perfect-gyrator-cond}).
The overall phase varies continuously with $r$ from $\pi/2$ to $-\pi/2$, attaining the maximum and minimum value for the values of $r$ corresponding to $\alpha=0$, and it is zero for the value of $r$ corresponding to $\alpha=1$.

We focus on the small $\alpha$ regime, with $\varphi\approx \pm \pi/2$. From Eq. (\ref{eq:circulation-condition}), it easily follows that by choosing a combination of $\pi/2$ out-of phase directional couplers, with $\beta=-1$, good circulation is achieved with the minimal possible size, i.e. $L_B\approx 0$.
In this situation, the physical scale of the circulator is determined by the gyrator and by the directional couplers. Not requiring impedance matching, the three-terminal gyrator can be made quite compact and the main limitation on the scalability is set by the directional couplers, which typically rely on ring $\lambda/4$ resonators.
Although in the microwave regime the wavelength $\lambda$ is typically few centimeters, the directional couplers can be miniaturized to $\lambda/N$, $N\approx 20$, by standard microwave engineering tricks \cite{Simons}.

We now quantitatively study the performance of the QH circulator, introducing the parameters \cite{Viola-DiVincenzo, Placke}
\begin{subequations}
\label{eq:Q_cw_acw}
\begin{flalign}
    Q_{\circlearrowleft} &\equiv \lvert S_{12} \rvert + \lvert S_{23} \rvert + \lvert S_{31} \rvert \leq 3, \\
    Q_{\circlearrowright} &\equiv \lvert S_{21} \rvert + \lvert S_{32} \rvert + \lvert S_{13} \rvert \leq 3,
\end{flalign}
\end{subequations} 
where the equality corresponds to perfect circulation in the direction of the arrow.

\begin{figure}
\includegraphics[scale=0.6]{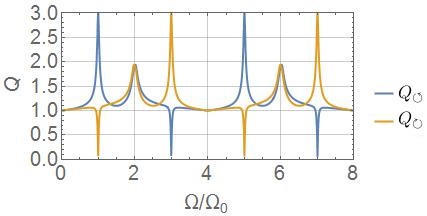}
\caption{$Q_{\circlearrowleft}$ and $Q_{\circlearrowright}$ as a function of $\Omega/\Omega_0$ in most compact scenario possible, with $L_A=L_B=L_C=L_D=0$. We used the self-matched HE gyrator with $r=2$ and $\alpha=0.04$ and the directional couplers are $\pi/2$ out-of-phase, i.e $\beta=-1$.}
\label{fig:Qa-Qc}
\end{figure}

We focus the analysis on the case $r=2$, which guarantees both self-matching and tolerance against parasitics as discussed in Sec. \ref{sec:gyrator}.
Fig. \ref{fig:Qa-Qc} shows the plots of the $Q$ parameters as a function of frequency when $\beta=-1$, $L_B=0$ and $\alpha=0.04$. It shows that excellent circulation can be attained in either direction, depending on the gyration frequency chosen.
A small remark is in order here: in our model, we have assumed all-pass directional couplers, which is not the case for on-chip devices. However, $\lambda/4$-resonators have a bandwidth typically much larger than the one of the gyrator, which consequently is the main limiting factor on the frequency performances.

\section{Conclusion}
In the present study we analyze an alternative set-up for a Hall-effect gyrator.
The implementation we present is experimentally easier to fabricate with respect to the previous proposal by VD, and it exhibits a very interesting self-matching property. We predict also that this device will be, to good approximation, insensitive to parasitic coupling between electrodes.
Finally, we incorporate this construction into an interferometer and we verify that this set-up behaves as a well-functioning circulator in an acceptable range of frequencies, and that its minimal size is mainly limited by the size of the two directional couplers.

\section{Acknowledgments}
The authors would like to thank D. Reilly, A. Mahoney, A.C. Doherty, G. Verbiest, C. Stampfer and L. Banszerus for useful discussions.
This work was supported by the Alexander von Humboldt foundation.

\bibliography{lit}

\begin{appendix}
\section{\label{appendix:S-parameter}Scattering parameters}
We report here the explicit expressions for the $S$-matrix elements in Eq. (\ref{eq:s-g-matrix}):
\begin{align}
S_{11} &= \frac{g(\Omega,r,\alpha)}{f(\Omega,r,\alpha)}, \\
S_{21} &=-\frac{4 \alpha  \left(1-e^{i \Omega }\right)^2}{f(\Omega,r,\alpha)},
\end{align}
with 
\begin{multline}
f(\Omega,r,\alpha)\equiv -\alpha ^2 e^{2 i \Omega }+2 \alpha  (\alpha +2) e^{i \Omega }-(\alpha +2)^2+\alpha ^2 e^{i r \Omega }+\\(\alpha -2)^2 e^{i (r+2) \Omega }-2 \alpha  (\alpha -2) e^{i (r+1) \Omega },
\end{multline}
and 
\begin{multline}
g(\Omega,r,\alpha)\equiv \alpha ^2 \left(-e^{i r \Omega }+2 e^{i (r+1) \Omega }-2 e^{i \Omega }+e^{2 i \Omega }\right)+\\
\left(\alpha ^2-4\right)\left( 1-e^{i (r+2) \Omega }\right).
\end{multline}
\end{appendix}

\end{document}